**TITLE:**

Intuitive Neuromyoelectric Control of a Dexterous Bionic Arm Using a Modified Kalman Filter

**AUTHORS:**

Jacob A. George[a], Tyler S. Davis[b], Mark R. Brinton[a], Gregory A. Clark[a]

[a]Department of Biomedical Engineering, University of Utah, Salt Lake City, Utah, 84112, United States

[b]Department of Neurosurgery, University of Utah, Salt Lake City, Utah, 84112, United States

Jacob.George@utah.edu, Tyler.Davis@hsc.utah.edu, Mark.Brinton@utah.edu, Greg.Clark@utah.edu

**CORRESPONDING AUTHOR:** Jacob A. George (Jacob.George@utah.edu)

**ABSTRACT:**

**Background:** Multi-articulate prostheses are capable of performing dexterous hand movements. However, clinically available control strategies fail to provide users with intuitive, independent and proportional control over multiple degrees of freedom (DOFs) in real-time.
**New Method:** We detail the use of a modified Kalman filter (MKF) to provide intuitive, independent and proportional control over six-DOF prostheses such as the DEKA "LUKE" Arm. Input features include neural firing rates recorded from Utah Slanted Electrode Arrays and mean absolute value of intramuscular electromyographic (EMG) recordings. Ad-hoc modifications include thresholds and non-unity gains on the output of a Kalman filter.
**Results:** We demonstrate that both neural and EMG data can be combined effectively. We also highlight that modifications can be optimized to significantly improve performance relative to an unmodified Kalman filter. Thresholds significantly reduced unintended movement and promoted more independent control of the different DOFs. Gain were significantly greater than one and served to amplify participant effort. Optimal modifications can be determined quickly offline and translate to functional improvements online. Using a portable take-home system, participants performed various activities of daily living.
**Comparison with Existing Methods:** In contrast to pattern recognition, the MKF allows users to continuously modulate their force output, which is critical for fine dexterity. The MKF is also computationally efficient and can be trained in less than five minutes.
**Conclusions:** The MKF can be used to explore the functional and psychological benefits associated with long-term, at-home control of dexterous prosthetic hands.

**HIGHLIGHTS:**

- Neural and electromyographic signals can be used together to estimate motor intent
- Ad-hoc thresholds and gains improve prosthetic control by reducing unintended movement
- Thresholds and gains can be optimized offline and translate to functional improvements online

- Computational efficiency enables training and real-time use of a portable prosthetic control system in a home environment
- Participants successfully performed various activities of daily living in lab and at home

**ABBREVIATIONS:**

Utah Slanted Electrode Array (USEA); Intramuscular Electromyographic Recording Leads (iEMGs); Electromyography (EMG); Degree of Freedom (DOF); Kalman Filter (KF); Modified Kalman Filter (MKF); Optimal Kalman Filter (OKF).

## 1. INTRODUCTION:

Multi-articulate prosthetic hands are now capable of reproducing the complex movements of endogenous human hands [1]–[3]. However, users' ability to intuitively control these advanced prostheses is still limited, and current prosthetic control remains unsatisfactory [4], [5]. Despite the fact that the 6+ degrees of freedom (DOFs) on these prostheses can move independently, the majority of control algorithms utilize classification-based approaches where users select between pre-determined grip patterns with a fixed force output [2], [6], [7]. There have been only limited attempts to provide users with independent and proportional control (i.e., simultaneous regression), and these studies all fall short of reaching the number of DOFs available on currently available multi-articulate prosthetic hands.

The use of classifiers, or regression of a smaller number of DOFs, for prosthetic control is due in part to the limited availability of intuitive control signals. Control of the most dexterous prostheses currently on the market relies on a series of gyroscopic inputs from the residual limb or intact feet [2], [7]. Arguably, the most intuitive control signals available to amputees are those originating from their residual nerves and/or muscles. In this work, we use a modified Kalman filter (MKF) to extract motor intent from the neural and/or electromyographic (EMG) signals that persist after a transradial amputation, and demonstrate that the MKF improves performance relative to a traditional, unmodified Kalman filter (KF).

There is a strong mathematical and theoretical justification for using a KF to estimate motor intent [8]. KFs have been used for decoding motor intent previously [9], [10]. Our group has also previously reported on the use of a KF for estimating motor intent from neural recordings offline [11], [12]. However, in online tasks that involve real-time control with visual and/or sensory feedback, participants expressed a desire to make the control more consistent and responsive. To this end, we introduced simple ad-hoc modifications to the KF that subjectively improved the participants' control [13]–[16]. Until now however, there was no objective measure of how these modifications impact performance. This question is particularly intriguing given that patient preference for ad-hoc modifications deviates from the mathematical and theoretical basis of the KF.

In the present report, we first demonstrate that neural recordings and EMG recordings can be used together as inputs to a KF. Then, we show that that modifying the output of the KF significantly improves performance on objective and clinically relevant measures. We further validate the MKF by demonstrating successful completion of activities of daily living (ADLs) when controlling a six-DOF prosthesis – first in a lab environment, and then under supervision at a participant's home, using a portable take-home system. This pilot take-home study highlights the technological feasibility and the practical benefits of the MKF as a clinical tool for amputees. To our knowledge, this is the first implementation of a portable neuromyoelectric prosthetic system providing independent and proportional control of six DOFs.

## 2. METHODS:

### 2.1. Human subjects

We implanted Utah Slanted Electrode Arrays (USEAs; Blackrock Microsystems, Salt Lake City, UT, USA) and intramuscular EMG recordings leads (iEMGs; Ripple LLC, Salt Lake City, UT, USA) into two transradial amputee participants (Fig. 1A). The first amputee (subject 6; S6) was a 57-year-old, left-hand-dominant male whose left foot and left forearm were amputated 13 years prior, after an electrocution injury [13]. The second amputee (subject 7; S7) was a 48-year-old, left-hand-dominant male, whose right forearm was amputated after Complex Regional Pain Syndrome (CRPS) and multi-year disuse [14]. Our work with five prior subjects (S1 – S5) have been previously published [11], [12], [15]. We also recruited six intact participants for surface electromyographic (sEMG) experiments to further validate our approach with the amputees. Two of the intact subjects with prior myoelectric experience were co-authors on this study. Details regarding the participants involved in this study are listed in Table 1.

Table 1: Amputee and Intact Participants

| Participant ID | Age | Sex | Dominant Hand (prior to amputation) | Forearm Tested | Amputation | Prior Experience |
|---|---|---|---|---|---|---|
| S6 | 57 | Male | Left | Left | 13 *years* prior | Owns myoelectric prosthesis |
| S7 | 48 | Male | Left | Right | 3 *weeks* prior | None |
| Intact 1 | 35 | Male | Right | Left | None | Prior sEMG experiments |
| Intact 2 | 41 | Male | Left | Left | None | Prior sEMG experiments |
| Intact 3 | 23 | Female | Right | Right | None | None |
| Intact 4 | 22 | Male | Left | Right | None | None |
| Intact 5 | 22 | Male | Right | Right | None | None |
| Intact 6 | 38 | Male | Right | Left | None | None |

Both amputee participants (S6 and S7) were evaluated by a physician and a psychologist for their willingness and ability to participate in the study. All surgeries and experiments were performed with informed consent from the participants and following protocols approved by the University of Utah Institutional Review Board and the Department of Navy Human Resources Protection Program.

### 2.2. Neuromyoelectric devices

We implanted two USEAs in participant S6 (one in the median nerve and one in the ulnar nerve) and three USEAs in participant S7 (two in the median nerve and one in the ulnar nerve) (Fig. 1B). USEAs are silicon microelectrode arrays, with 100 electrode shafts, arranged in a 10 x 10 grid on a 4-mm x 4-mm base. Electrode shafts are spaced 400 μm apart, with shaft lengths varying from approximately 0.75–1.5 mm [17]. Four looped platinum wires were also implanted—two served as electrical ground and stimulation return, and two served as reference wires for recording. Four electrodes from the longest row of electrode shafts on the USEA were also sometimes used as an on-array electrical reference for recordings [18], [19]. USEAs electrodes were metalized with SIROF-deposited iridium oxide. The USEAs implanted in both subjects featured helical wire bundles with the helix extended all way into the connector system in S7 to further reduce wire breakage [20].

For both S6 and S7, we also implanted eight iEMG leads (with four contacts each), with attempted targeting of each lead to different lower-arm extensor or flexor muscles (Fig. 1C). A separate iEMG lead was implanted proximal and posterior to the elbow to provide an electrical reference and ground. Both USEA and iEMG electrodes exited the arm through a percutaneous incision and were mated with external Gator Connectors (Ripple LLC, Salt Lake City, UT, USA) and associated Front Ends (Ripple LLC, Salt Lake City, UT, USA) for filtering and amplifying the signal (see Signal Acquisition section below).

For sEMG validation experiments, we used custom-made neoprene sleeves embedded with 34 dry electrodes, one of which served as an electrical reference and another which served as an electrical ground. These custom sleeves were placed on the upper forearm of the intact participants, such that the 32 active electrodes were placed over the extrinsic hand muscles and the reference and ground electrodes were placed over the ulna, a few centimeters distal to the elbow.

### 2.3. Surgical procedure

Starting the day before the implant surgery, subjects were given an oral prophylactic antibiotic for seven days (100 mg minocycline, twice per day), which has been reported to improve neuronal recording quality [21]. Under general anesthesia, the USEAs were placed in median and ulnar nerves in the upper arm, several centimeters proximal to the medial epicondyle. The iEMGs were implanted midway along the forearm. The epineurium was dissected and USEAs were implanted using a pneumatic inserter tool [22]. After insertion, the epineurium was sutured back together around each USEA (if there was enough epineurium available), and collagen wraps (AxoGuard Nerve Protector, Axogen, Alachua, FL, USA) were secured around the nerve and the ground and reference wires with vascular clips. A 0.1 mg/kg dose of dexamethasone was administered after tourniquet removal, which has been reported to reduce the foreign body response and improve neural recordings [23], [24].

The percutaneous wire sites were dressed using an antibiotic wound patch (Biopatch, Ethicon US LLC, Somerville, NJ, USA) and replaced at least once a month. The USEAs and iEMGs were surgically removed after 14 months for S6 and after 16 months for S7, as the result of prior mutual agreements between the volunteer participants and the experimenters regarding study duration. Both USEAs and iEMGs remained viable throughout the entire duration of the studies.

### 2.4. Signal acquisition

Neural and EMG recordings were sampled at 30 kHz and 1 kHz, respectively, using the Grapevine or Nomad Systems (Ripple LLC, Salt Lake City, UT, USA) (Fig 1D). Although three USEAs were implanted into S7, only two were used at a time due to software limitations. Continuous neural signals were band-pass filtered with cutoff frequencies of 0.3 Hz (first-order high-pass Butterworth filter) and 7500 Hz (third-order low-pass Butterworth filter). A digital high-pass filter with a cutoff frequency of 250 Hz (fourth-order Butterworth filter) was also applied and multi-unit activity was detected by threshold crossings of an adaptive, automated threshold, set to approximately negative six times the root mean square of the signal [12]. The unsorted multi-unit neural activity was binned into 33-millisecond windows and converted into firing rates. The firing rate was smoothed using an overlapping 300-ms window. The resulting neural feature set consisted of the 300-ms smoothed firing rate on 192 channels (96 active electrodes for each of the two USEAs), calculated at 30 Hz.

Continuous EMG signals (32 channels) were band-pass filtered with cutoff frequencies of 15 Hz (sixth-order high-pass Butterworth filter) and 375 Hz (second-order low-pass Butterworth filter). Notch filters were also applied at 60, 120, and 180 Hz. Differential EMG signals were calculated for all possible pairs of channels, resulting in 496 (32 choose 2) differential pair recordings. The mean absolute value (MAV) was then calculated for the 32 single-ended recordings and the 496 differential recordings at 30 Hz. The MAV was smoothed using an overlapping 300-ms window. The resulting EMG feature set consisted of the 300-ms smoothed MAV on 528 channels (32 single-ended channels and 496 differential pairs), calculated at 30 Hz.

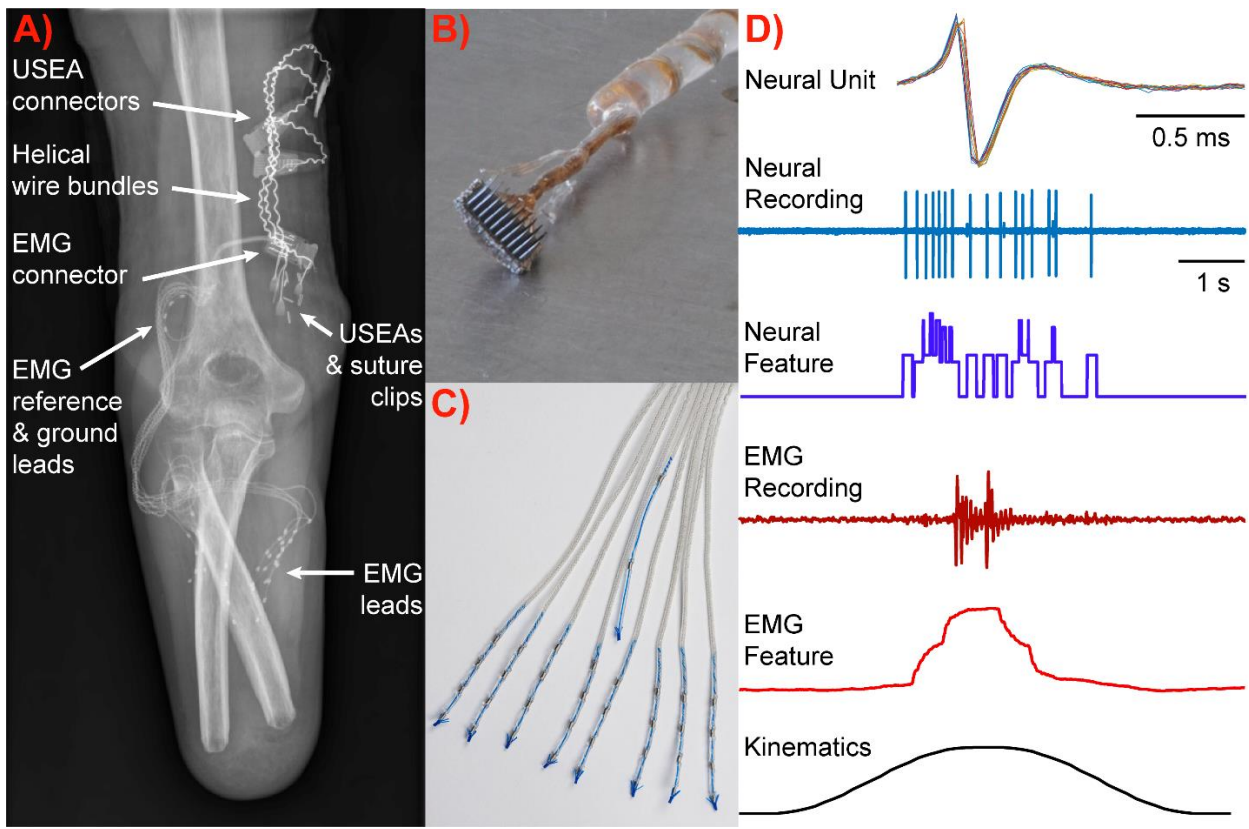

Figure 1: A) Two transradial amputees were implanted with Utah Slanted Electrode Arrays (USEAs) and electromyographic (EMG) recording leads. (B) USEAs were implanted into the residual median and ulnar nerves. C) EMG leads were implanted into the residual forearm muscles. D) Example neural and EMG data recorded simultaneous while amputee participant S7 actively mimicked a middle finger flexion queue. Overlaid neural units (Neural Unit) from 16 sequential spike events (Neural Recording) demonstrate the robust recording capabilities of the USEA at 17 months after implantation. 33-ms-binned neural firing rates (Neural Feature) and mean absolute values of the EMG (EMG Feature) are used as inputs to a modified Kalman filter to estimate the amputees' motor intent (Kinematics).

### 2.5. Training datasets

In order to correlate neuromyoelectric signals to intended hand movements, we instructed the participants to mimic preprogrammed hand movements performed by a prosthetic or virtual hand. As the participants mimicked the hand movements, we recorded, in synchrony, the kinematics of the hand, EMG activity, and neural activity (Video S1). For example, the index finger of the prosthetic hand would flex, and the participant would actively attempt to flex their intact/phantom index finger simultaneously. In order to account for the participants' reaction time in mimicking the hand movements, we shifted the recorded kinematics by a lag that was determined by cross-correlating the kinematic and neuromyoelectric recordings [14].

In general, training datasets consisted of participant mimicking individuated movements: flexion/extension and abduction/adduction of the thumb (D1); flexion/extension of the index finger (D2); flexion/extension of the middle finger (D3); and flexion/extension and pronation/supination of the wrist. Movements were completed sequentially, such that all trials of a single movement were performed one after another.

#### 2.5.1 Virtual Hand-matching Task

For the virtual hand-matching tasks, the participants mimicked the movements of a virtual prosthetic hand (MSMS; Johns Hopkins Applied Physics Lab, Baltimore, MD) (Fig. 2). Virtual hand movements included individuated movements of each DOF (flexions/extensions of D1, D2, and D3; wrist flexion/extension; wrist pronation/supination; D1 abduction/adduction) as well as one combination movement (simultaneous extension of D1 and D2), for a total of 13 movements.

Each of the 13 movements was repeated four times. The total duration of each movement was 6.4 s (consisting of a 0.7 s flexion/extension away from the resting hand position, a 5-s hold-time at the maximum distance away from the resting hand position, and then a 0.7-s extension/flexion returning to the resting hand position). This duration of movements was shown to improve the ability to estimate motor intent using both linear and non-linear algorithms [14].

#### 2.5.2 Activities of Daily Living

Training datasets in preparation for ADLs involved the amputees' mimicking the movements of a physical prosthetic hand ("LUKE" Arm; DEKA, Manchester, NH). These movements included individuated movements of each DOF of the prosthesis (flexions/extensions of D1, D2, and D3; wrist flexion/extension; wrist pronation/supination; D1 abduction/adduction). Occasionally, two or three combination movements were included as well (e.g., simultaneous flexion of D1-D3; simultaneous extension of D1-D3; simultaneous flexion of D1 and D2).

For S6, each movement was performed ten times and the total duration of the movements was 1.5 s (consisting of a 0.7-s flexion/extension away from the resting hand position, a 0.1-s hold-time at the maximum distance away from the resting hand position, and then a 0.7-s extension/flexion returning to the resting hand position). For S7, number and duration of movements was changed in order minimize training time and improve performance (Fig. S1). For S7, each movement was performed four times and the total duration of the movements was 4.4 or 6.4 s (same paradigm described above, with either a 2-s hold-time or a 5-s hold-time).

*2.6. Kalman filter*

We used a KF, as defined in [8], [25], [26], in order to estimate motor intent from the continuous neural and/or EMG signals. Ad-hoc modifications to this KF are described later in section 2.8. The KF has been used by our group before [11], [12], and the detailed mathematical justifications, construction and parameters of the KF have been outlined in [8]. The baseline firing rate or MAV for each channel was subtracted from the features prior to training and testing of the KF. We assumed the neural and EMG features were normally distributed, and relied on the KF covariance matrix to inherently address differences between them. The combined neural and EMG feature set (720 channels) was reduced to 48 channels using a stepwise Gram-Schmidt channel-selection algorithm [27]. A single KF was used to predict all 6 DOFs of the prosthetic hand simultaneously. We limited the predictions of the KF between -1 and 1: -1 corresponded to maximum extension/adduction/supination, +1 corresponded to maximum flexion/abduction/pronation, and the hand was at rest at zero [12].

*2.7. Offline analysis of neural and EMG features*

We analyzed 66 training datasets from S6 and 22 training datasets from S7, collected over 14- and 17-month timeframes respectively. For each dataset, we used a random 50% of the trials for each DOF to train three different MKFs: one using neural data alone, another using EMG data alone, and a third using neural and EMG data combined. Using the remaining 50% of the data, we then compared the performance of the three MKFs.

Performance was measured by the root mean squared error (RMSE) between the MKF output and the true kinematics the participant was actively mimicking. We divided the RMSE into two categories: intended movement RMSE and unintended movement RMSE. Intended movement RMSE measures the ability to replicate the participants desired movements (e.g., flexion or extension of a single DOF), whereas unintended movement RMSE measures the ability to eliminate cross-talk such that only the intended DOF is active (Fig. 2).

For each participant, we used separate one-way non-parametric ANOVAs (Kruskal-Wallis) to compare the three MKFs (neural data, EMG data, and neural + EMG data) for intended movement RMSE and for unintended movement RMSE. If any significance was found, subsequent pairwise comparisons were made using the Dunn & Sidák correction for multiple comparisons.

*2.8. Modified Kalman filter*

Previous publications have mentioned the use ad-hoc thresholds and non-uniform gains on the output of the KF to improve performance [13], [14]. The effect of these modifications can be written as

$$Modified\ Output = \begin{cases} \dfrac{(Output\ \cdot\ Gain) - Threshold}{1 - Threshold}, & Output \geq Threshold \\ 0, & Output < Threshold \end{cases}$$

for the positive direction for each DOF (e.g., flexion/abduction/pronation). A similar formula is used for the negative direction that preserves the appropriate sign and directionality. Importantly, the gain is applied before the threshold in order to ease movement initiation. The ad-hoc gain described here, and throughout the manuscript, is distinctly different from the Kalman gain matrix that's generated during the construction of the KF.

Our previous publications used default values of 0.2 and 1 for the thresholds and gains respectively [13], [14], although they were often subsequently tuned on an individual DOF basis to provide optimal control [13]. Individual tuning generally involved modifying thresholds first, and then gains. Threshold were generally increased from 0.2 until unintended movement and baseline jitter were eliminated. Gains were then increased from 1 until the participant felt they could control the DOF comfortably. The final modified output, updated at 30 Hz, serves as scaled estimate of motor intent that emphasizes non-zero activity to make movements more distinct.

The KF is a recursive algorithm, and modifications would generally lead to an unstable output. To avoid this, we did *not* use the modified output as the recursive input to the KF, because this would cause a mismatch between the predicted state and the actual commanded state. Furthermore, to account for the non-uniform gains, the recursive input to the KF was bounded by the inverse of the gain on a DOF-by-DOF basis. The modified output was used directly for real-time position or velocity control of virtual or physical prosthetic hands.

### *2.9. Offline optimization and analysis of modifications*

To determine the benefit of these modifications on the KF, we analyzed the same 66 training datasets from S6 and 22 training datasets from S7 described above. Similarly, for each dataset, we used a random 50% of the trials for each DOF to train a KF. Using the remaining 50% of the data, we then compared the performance of the (unmodified) KF [11], [12], with the performance of an MKF used default parameters [13], [14], and with the performance of a MKF that used optimized parameters (optimized Kalman filter, OKF).

We found optimal thresholds and then optimal gains (both on a DOF-by-DOF basis) for each dataset using a sequential golden section search and parabolic interpolation optimization algorithm [28], [29]. We defined the optimal value as the value that results in the lowest root mean squared error (RMSE) for both intended movement and unintended movement, which were weighted equally using a simple linear model. We optimized the mean squared error, as opposed to variance, on a DOF-by-DOF basis based on the assumption that the full kinematic range would be utilized for each DOF.

For example, the optimization explored various threshold values using a gradient-based approach, training a KF on the first 50% of the data with that threshold value, and then testing the performance (RMSE) on the second half of the data. After an optimal threshold was identified, the process was repeated to determine the optimal gain – training a KF on the first 50% of the data with the optimal threshold value and that gain, and then testing the performance (RMSE) on the second half of the data.

For each participant, we used separate one-way non-parametric ANOVAs (Kruskal-Wallis) to compare the three KFs (unmodified, modified, and optimally modified) for intended movement RMSE and for unintended movement RMSE. If any significance was found, subsequent pairwise comparisons (Wilcoxon rank sum tests) were made using the Dunn & Sidák correction for multiple comparisons.

### *2.10. Real-time virtual hand-matching task*

*2.10.1 Task summary*

For S7 and six intact participants, we quantified the benefit of thresholding the output of the KF in a virtual hand-matching task where the participant had real-time visual feedback [14]. We focused solely on the threshold modification because the thresholds occur outside of the KF model, and thus can be updated in real-time without retraining the KF.

In this task, the participants actively controlled the virtual MSMS hand and attempted to move select DOF(s) to a target location (Fig. 2). Target locations were at 50% of the maximum amount of flexion/extension possible in order to evaluate proportional control. The participant was instructed to hold the selected DOF(s) at the target location for the trial duration. Visual feedback was provided to the participant to confirm when each DOF was within ±5% of target location.

Each test trial lasted 7 s, and there was a 10-s wait time between trials to avoid fatigue for the amputee (1 s for intact individuals). A total of four trials were tested for 16 different movements. These movements included the 13 movements that were part of the original training dataset, and an additional three novel, untrained movements (simultaneous flexion/extension of D1, D2 and D3; and simultaneous flexion of D1 and D2). These novel combination movements are not a pure linear summation of the simpler individual movements; thus, they provide a test of decode generalizability to clinically relevant, untrained movements.

For each participant, we completed the virtual hand-matching task with an unmodified KF and a MKF in a single experimental session. The KF and MKF were blinded from the participants and appeared in a randomized, counterbalanced order. We first collected a training dataset for each participant as described in the previous section. Using a random 50% of the training dataset, we trained a KF, and then, using the remaining 50% of the training dataset, we determined an optimal threshold value applied uniformly across all DOFs. A single optimal threshold was applied across all DOFs in order to reduce computational time.

### 2.10.2 Performance metrics

We analyzed both offline performance (based on the training data) and online performance (based on the participant's performance during the extend hand-matching task).

For offline analysis, we compared the intended and unintended movement RMSE of the unmodified KF and the MKF (with an optimal threshold value) on the 50% of the dataset that was not used to train the KF.

For online analysis, we compared the performance of the two algorithms on the virtual hand-matching task using three metrics: 1) intended movement RMSE; 2) unintended movement RMSE (i.e., cross-talk), and 3) the mean longest continuous-hold duration (i.e., hold duration) within the desired 10%-error window around the target location [14]. To calculate RMSE during intended and unintended movement, devoid of the participant's reaction time, we delayed the recorded kinematics by a lag that was determined by cross-correlating the kinematic estimate and target location signals. This alignment was applied across all experimental conditions for a given session, so that there would be no bias affecting one experimental condition more than another.

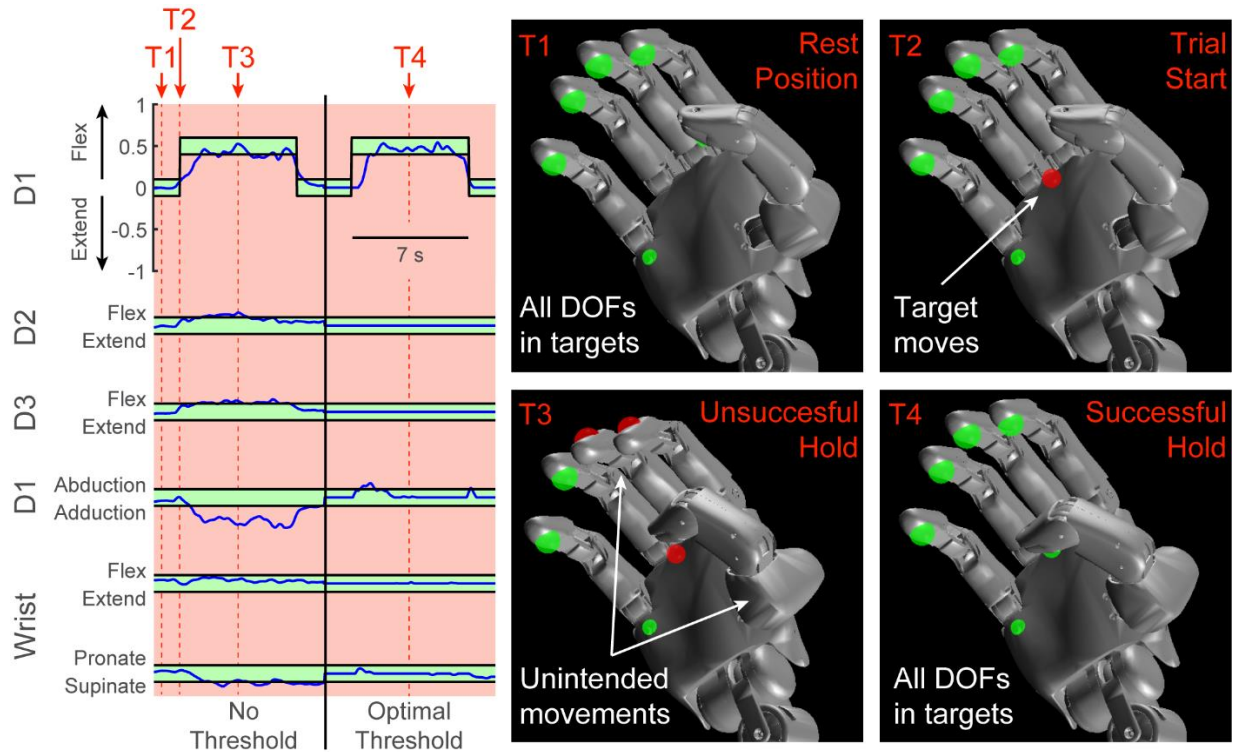

Figure 2: Real-time virtual hand-matching task. The participants had real-time proportional control over a six-degree-of-freedom (DOF) virtual prosthetic hand. Blue traces show the output of the Kalman filter (unmodified, left; modified with optimal threshold, right) for each DOF. The participant is instructed to keep each DOF within the 10%-error window for the target locations (green regions, left; shown to participant in real-time with green spheres, right). When outside of the 10%-error window (red regions, left) the participant is notified in real-time by the red sphere on each DOF (right). Timepoints A–D (top left) highlight important parts of the task and have corresponding images on the right. A) Before the start of the trial, the participant is relaxed and all six DOFs are within a 10%-error window from their resting positions. B) At the start of each trial, a selected DOF(s) moves to a target location (e.g., D1 moves to 50% flexion) and the participant must actively control the DOF(s) to match the desired location. C) Although D1 is within a 10%-error window from the target location (i.e., low RMSE for the intended movement), the other DOFs are not stationary (i.e., high RMSE for unintended movement). D) With optimal thresholds, the participants were better able to keep all six DOFs within their target locations. The total duration that all DOFs were within their target locations is defined as the hold duration – in the example shown, this value is approximately 0 s for the no-threshold condition and approximately 3.6 s for the optimal-threshold condition.

*2.10.3 Statistical analysis*

Statistical analyses were performed separately for the amputee and the intact subjects, due to existing and potential differences between these two groups (e.g., neural data and intramuscular EMG recordings for implanted amputee subjects only). Statistical analyses were also performed separately for intended and unintended movement RMSE, due to the different nature of these two metrics. Outliers (more than 1.5 times the interquartile range (IQR) above the upper quartile or below the lower quartile) were removed prior to inferential statistical analyses.

For the amputee, we performed an unpaired nonparametric t-test (Wilcoxon rank sum test) to compare the RMSEs (for all movements and trials) of the unmodified KF and the MKF (with an optimal threshold). Because these data were collected from a single amputee participant, we do not attempt to generalize these findings to the broader amputee population.

For the six intact participants, we performed a Friedman's test to compare the RMSEs (mean across movements, treating each trial as a separate observation) of the unmodified KF and the MKF. Because these data were collected from multiple participants and collapsed for each patient (i.e., $N = 6$), these findings can be generalized to a broader population of intact individuals with demographics similar to those reported in Table 1.

### *2.11. Real-world use in activities of daily living*

Participants S6 and S7 also completed various ADLs in the lab environment using the MKF with default thresholds and gains. For these tasks, the participants used the MKF to control a socket-mounted, six-DOF prosthetic hand and wrist – the DEKA "LUKE" Arm. The output of the MKF was used to directly control the position or velocity of the six DOFs of the prosthesis. The ability to proportionally control position or velocity was toggled on a DOF-by-DOF basis. In general, S6 preferred velocity control of all six DOFs, whereas S7 preferred position control for D1 – D3 and velocity control for the wrist DOFs.

Participant S7 also completed a preliminary supervised take-home trial which consisted of training the MKF and then performing various ALDs in his own home. iEMG recordings were used as inputs to the MKF, which was implemented on the portable battery-powered Nomad system (Ripple LLC, Salt Lake City, UT, USA) and communicated with the DEKA "LUKE" Arm with CAN communication (Fig. 3).

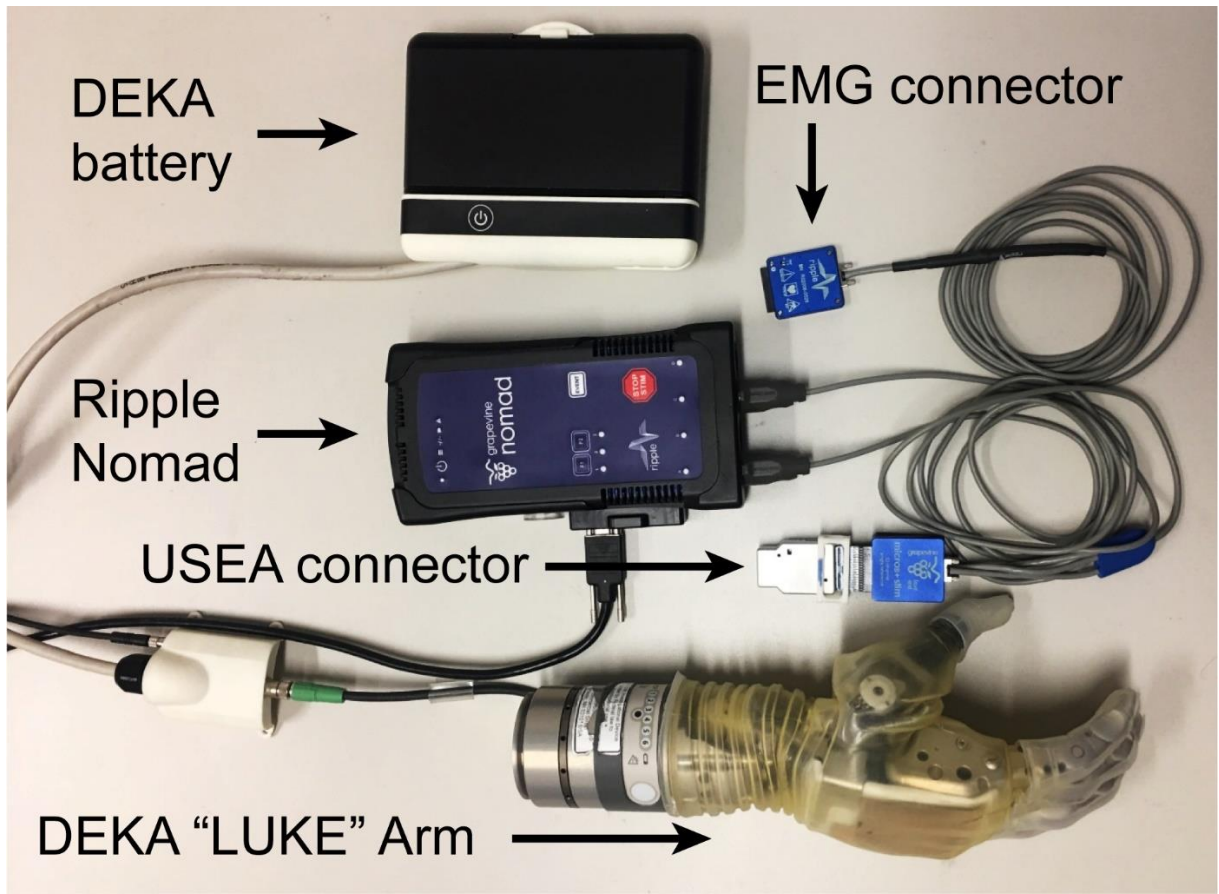

Figure 3: Portable system. The modified Kalman filter was deployed onto a portable processing unit (Ripple Nomad) that interfaces with implanted Utah Slanted Electrode Arrays (USEA) and intramuscular electromyographic recording leads (EMG) to control the DEKA "LUKE" Arm.

## 3. Results:

### 3.1. Neural and EMG signals can be combined for improved motor decoding

We implanted USEAs and EMG recording leads into the residual arm nerves and muscles of two transradial amputees. Using the MKF, we estimated user intent using both neural recordings and EMG recordings simultaneously. We demonstrate that for one amputee (S7), the combination of neural and EMG data resulted in the best performance ($p < 0.001$, rank sum tests; Fig. 4). For S7, a median of 8 and maximum of 17 neural channels were used as inputs to the MKF algorithm—respectively 17% (median) and 35% (maximum) of the 48 total channels used in the decode (Fig. 5). For S6, neural activity could no longer be recorded partway through the study – due to issues with older versions of the USEA, such as broken wires. Because no information was available through neural recordings, the best decode performance was achieved with myoelectric control alone (Fig. 4) – further indicated by the low number of neural channels utilized when both neural and EMG recordings were used as the input to the MKF (Fig. 5).

The Gram-Schmitt channel selection algorithm utilized to select the best 48 channels from the combined neural and EMG features is agnostic to channel type; the algorithm adds channels in a stepwise manner on the basis of how much channels are able to reduce estimation error. The large number of neural units selected here, particularly for S7 (Fig. 5), highlights the ability of the USEA to chronically record useful neural activity.

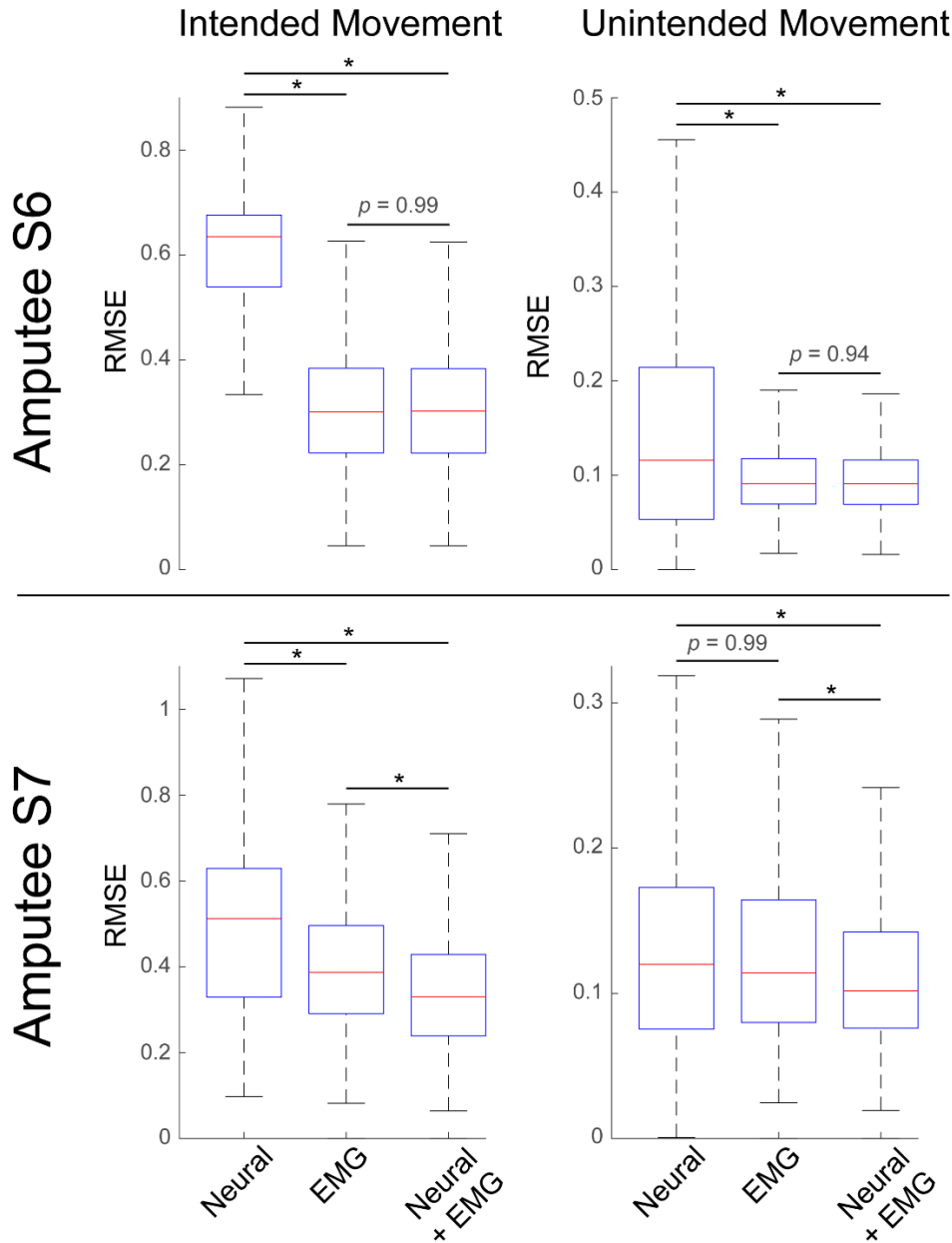

Figure 4: Motor intent can be decoded with a Kalman filter using neural and electromyographic (EMG) recordings, separately or combined. Neural firing rates were recorded from Utah Slanted Electrode Arrays implanted in residual arm nerves and the mean absolute value was calculated from EMG recording leads implanted in residual forearm muscles. Data show root mean square error (RMSE) of intended movement (Left) and unintended movement (Right) for 66 datasets from amputee participant S6 (Top) and 22 datasets from amputee participant S7 (Bottom). For S7, the best performance was achieved when neural and EMG features were used together. For S6, neural data provided no additional benefit over EMG data, due to limited recordings of neural activity. Boxplots show median (red line), IQR

(blue box) and most extreme, non-outlier values (black whiskers). * *p* < 0.001, Wilcoxon rank sum tests with Dunn & Sidák correction for multiple comparisons.

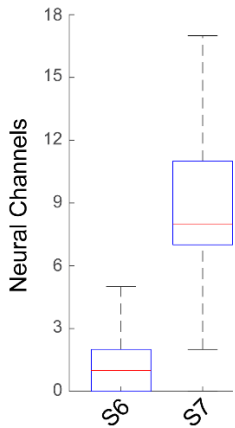

Figure 5: Utah Slanted Electrode Arrays implanted in residual arm nerves recorded multi-unit activity useful for estimating motor intent. A total of 48 neural or electromyographic channels were selected as input to a Kalman filter using a Gram-Schmidt channel-selection algorithm that adds features (agnostic to the source), stepwise based on how much they reduce estimation error [27]. For S7, a median of eight (17%) and a maximum of 17 (35%) neural channels were used as input to a Kalman filter. These numbers demonstrate that neural recordings are present and provide useful information throughout the 14- and 17-month implants. Boxplots show median (red line), IQR (blue box) and most extreme, non-outlier values (black whiskers).

### *3.2. Default thresholds reduce unintended movement but can affect intended movement (offline)*

We compared the performance of an unmodified KF and a MKF (with our traditional 20% threshold and uniform gain [13], [14]) across 66 datasets for S6 and 22 datasets for S7. We found that the added threshold (non-optimized default value of 0.2) significantly reduced unintended movement ($p < 0.001$, rank sum tests; Fig. 6). Specifically, the threshold reduced the median RMSE of unintended movement (i.e., cross-talk) by 90% for S6 and by 84% for S7. However, the added threshold also had a significant degradation on intended movement ($p < 0.001$, rank sum tests), resulting in a 15% increase in median RMSE for S6 and a 20% increase for S7.

### *3.3. Optimal gains and thresholds reduce unintended movement without degrading intended movement (offline)*

We previously used non-uniform gains as a method to counteract the negative effects of thresholds on intended movement [13], which could make movement more effortful. However, these gains were adjusted manually by the experimenter, on a DOF-by-DOF basis, using feedback from the participant. Here we demonstrate that the values for both thresholds and gains can be optimized to provide a significant reduction in unintended movement without any significant detriment to intended movement.

Optimizing thresholds and gains on a DOF-by-DOF basis improved subjects' performance compared with performance with the unmodified KF. We found that the OKF resulted in a significant reduction in the RMSE of unintended movement ($p$'s < 0.001, rank sum tests) – a 63% median reduction for S6 and a 47% median reduction for S7 – and no significant change in in RMSE of intended movement ($p$ = .43 and $p$ = 0.30, rank sum tests) (Fig. 6).

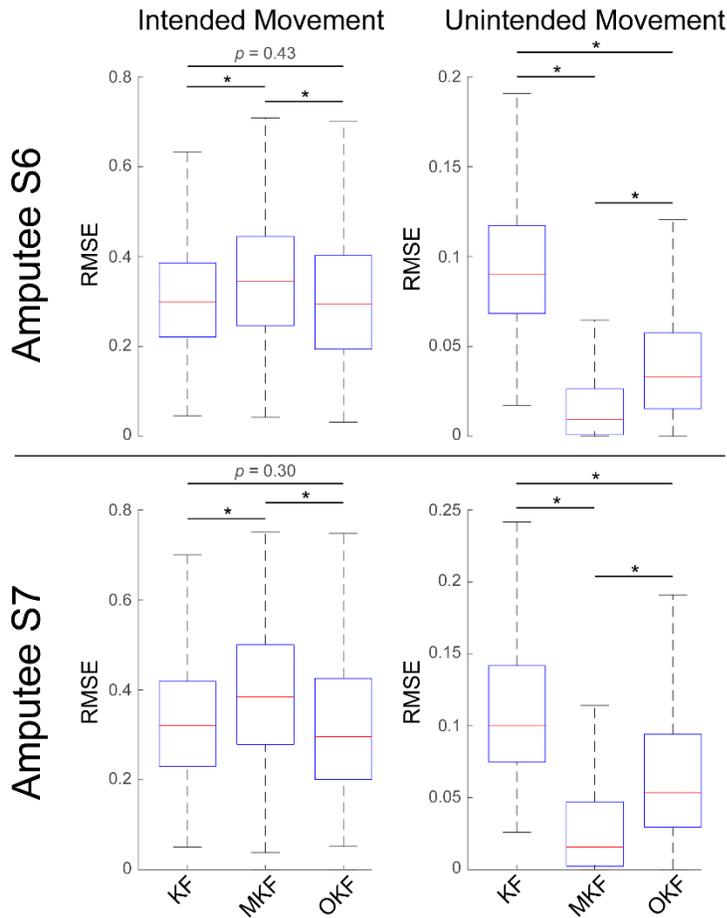

Figure 6: A modified Kalman filter that uses optimal values for thresholds and gains (OKF) provides improved performance relative to an unmodified Kalman filter (KF) and a modified Kalman filter that uses previously reported thresholds and gains (MKF). Data show root mean square error (RMSE) of intended movement (Left) and unintended movement (Right) for 66 datasets from amputee participant S6 (Top) and 22 datasets from amputee participant S7 (Bottom). The MKF produced significantly less unintended movement (i.e., cross-talk) than did either the KF or the OKF. However, the MKF also resulted in significantly worse performance on intended movement relative to performance with the KF or OKF. The OKF was comparable to the KF for intended movement (no significant differences), but still provided a significant reduction in unintended movement. Boxplots show median (red line), IQR (blue box) and most extreme, non-outlier values (black whiskers). * $p$ < 0.001, Wilcoxon rank sum tests with Dunn & Sidák correction for multiple comparisons.

### *3.4. Optimal modifications to the KF are DOF-specific and different from previously utilized values*

Optimal values for thresholds and gains were unique to each DOF and, in practice, non-zero thresholds were consistently utilized and gains were consistently non-uniform (i.e., not equal to one) (Fig. 7). Optimal thresholds were generally greater than 5%, but less than the 20% previously used as a default value [13], [14]. The median (± IQR) threshold across all DOFs was 0.15 ± 0.11 for S6 and 0.09 ± 0.12 for

S7. For both participants, median thresholds were significantly greater than 0 ($p < 0.001$ for both S6 and S7; Wilcoxon signed rank test). The median (± IQR) gain across all DOFs was 1.11 ± 0.18 for S6 and 1.23 ± 0.29 for S7. For both participants, median gains were significantly greater than 1 ($p < 0.001$ for both S6 and S7; Wilcoxon signed rank test), serving to counteract the thresholds by amplifying decode output and thereby easing the participants' effort.

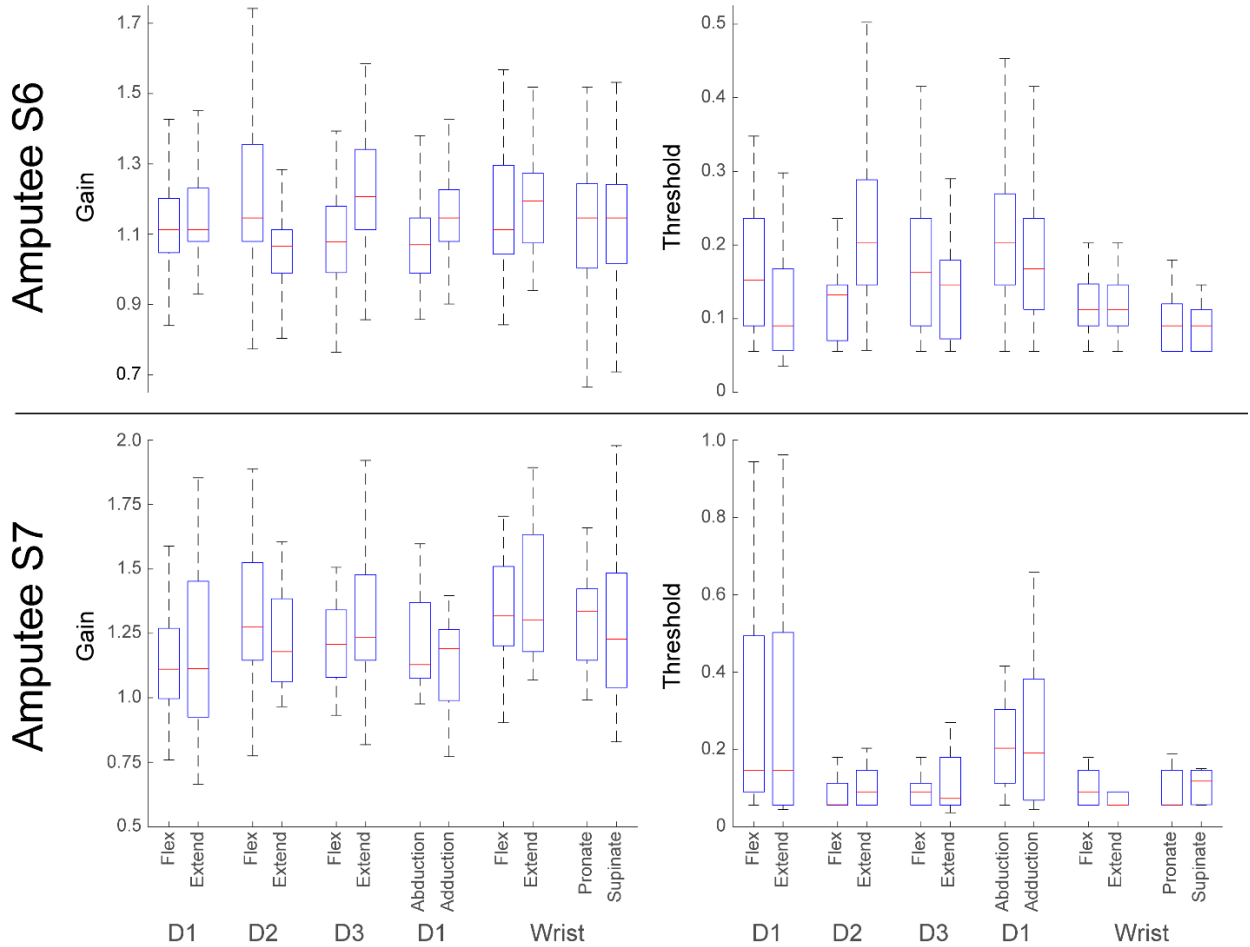

Figure 7: Distributions of optimal gains (Left) and thresholds (Right) for 66 datasets from amputee participant S6 (Top) and 22 datasets from amputee participant S7 (Bottom). For both participants, thresholds were significantly greater than zero ($p < 0.001$ for both S6 and S7; Wilcoxon signed rank test), but they were often less than previously used values [13], [14]. Non-zero thresholds were corrected using non-uniform gains, which were significantly greater than one ($p < 0.001$ for both S6 and S7; Wilcoxon signed rank test) – acting to amplify decode output and ease the participants' efforts. Boxplots show median (red line), IQR (blue box) and most extreme, non-outlier values (black whiskers).

### *3.5. Offline optimization translates to improvements in real-time tasks involving visual feedback (online)*

The above results demonstrate that the optimal modifications are specific to each DOF. However, optimizing 12 movements (two directions for 6 DOFs) is computationally expensive, and patient time is critical. We therefore sought to test the benefit of using an optimal threshold applied across all DOFs

(i.e., the same threshold applied to all movements). We calculated this optimal threshold for six intact participants and one amputee participant (S7) (Table 2). Similar to how we analyzed results presented above, we compared the performance of the unmodified KF with that of the MKF (with the optimal threshold applied across all DOFs) through an offline analysis. In addition, we also compared the performance of the two KFs online, in a virtual hand-matching task where the participants had visual feedback in real-time.

Table 2: Optimal threshold across all DOFs identified for each participant

| Participant ID | Optimal Threshold |
| --- | --- |
| S7 | 0.12 |
| Intact 1 | 0.13 |
| Intact 2 | 0.12 |
| Intact 3 | 0.13 |
| Intact 4 | 0.17 |
| Intact 5 | 0.16 |
| Intact 6 | 0.16 |
| **Mean Value** | **0.14** |

For both intact participants and S7, optimal thresholds had the same effect online as offline — the added threshold significantly reduced unintended movement in real-time when the participants had visual feedback ($p < 0.001$ for intact, Friedman's test; $p < 0.001$ for S7, rank sum test). Together, there were median RMSE reductions of 67% (offline) and 47% (online) for the intact participants, and 60% (offline) and 49% (online) for S7 (Fig. 8). Furthermore, the optimal threshold alone generally resulted in no significant difference in the median RMSE of intended movement, suggesting that gains may not be necessary for all improvements. These results demonstrate that thresholds optimized offline can predict improvements in performance online.

### 3.6. Modifications to the Kalman filter result in functional improvements

The above results demonstrate how offline optimization can be used to improve performance on online tasks. However, RMSE metrics are difficult to translate into tangible benefits, such as those you'd expect to see when performing ADLs with a physical prosthesis. To better address functional improvements, we quantified the longest duration that participants could hold all DOFs of the virtual hand within the 10%-error window used in the virtual hand-matching task. This metric—hold duration —provides a better estimate of the participant's ability to provide a precise and proportional grip over an object, similar to that needed to manipulate a fragile object.

With optimal thresholds, the intact participants achieved a 301% median increase in hold duration ($p < 0.001$, Friedman's test) (Fig. 8). Similarly, S7 also exhibited a significant improvement in hold duration ($p < 0.001$, rank sum tests) using an optimal threshold – although the difficulty of the task introduced a floor effect.

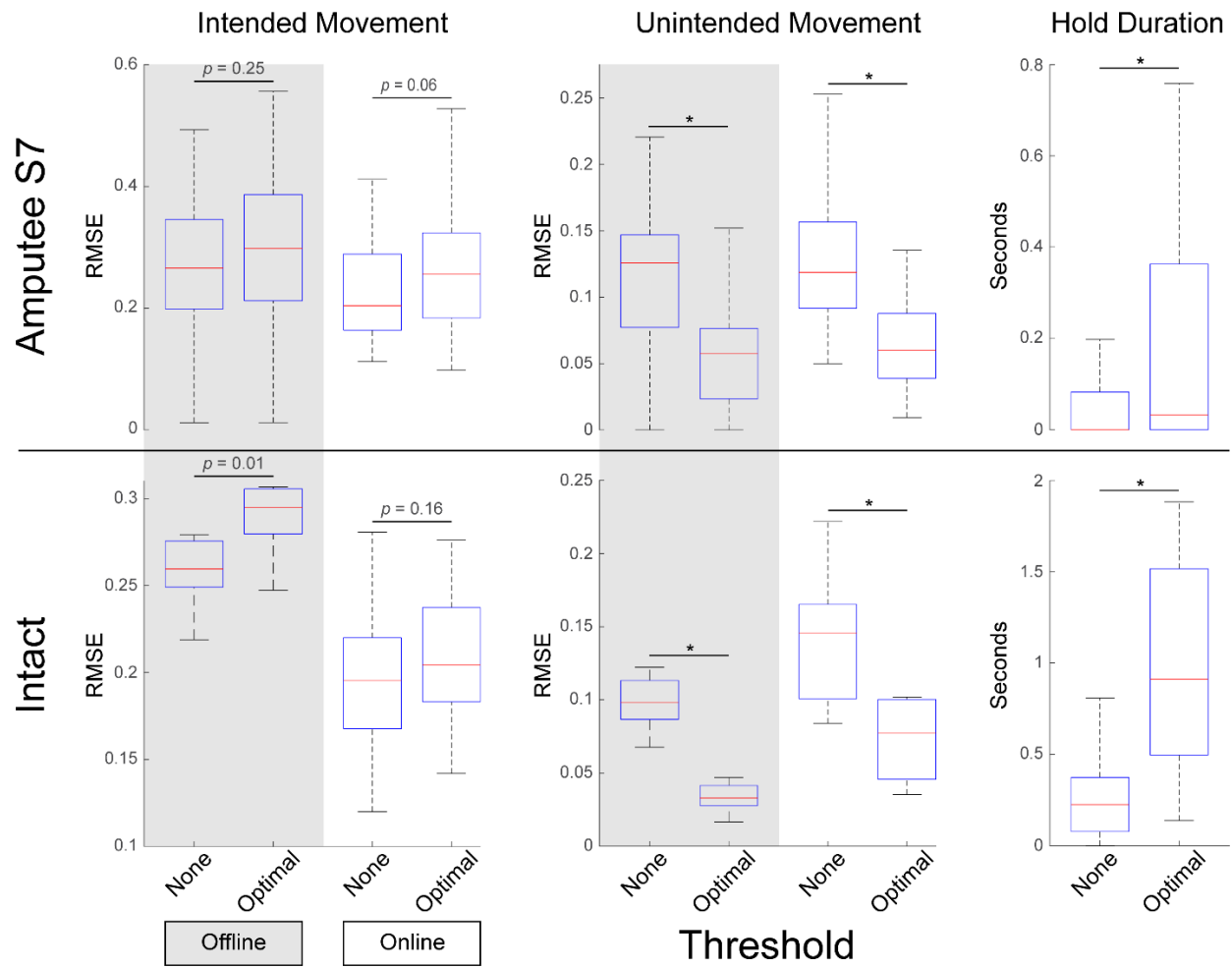

Figure 8. Optimal thresholds improve performance. Optimal thresholds were determined from a random 50% of the data collected from amputee participant S7 (Top) and six intact individuals (Bottom). When tested on the remaining 50% of the data, the added optimal thresholds improved performance (Offline, gray). These thresholds optimized offline – without the participant in the loop – were then used to improve performance in an online virtual hand-matching task – where the participant had visual feedback Online, white). For both offline and online tests, the optimal threshold had no impact (or a slightly negative impact) on the root mean squared error (RMSE) of intended movement (Left). The optimal threshold did however yield a significant reduction in unintended movement (Center). Together, this led to a significant improvement in a more functional metric of online performance – hold duration (Right). Boxplots show median (red line), IQR (blue box) and most extreme, non-outlier values (black whiskers). * $p < 0.001$, Wilcoxon rank sum tests for S7 and Friedman's tests for intact participants.

### 3.7. Participants performed various activities of daily living

Virtual tasks, such as the virtual hand-matching task, provide value insight into how an amputee will perform with a physical prosthesis [15]. To demonstrate control of multi-articulate prosthetic arms using the MKF, participants performed various ADLs using a 6-DOF prosthetic hand and wrist – the DEKA "LUKE" Arm. Participants performed both unimanual (prosthesis alone) and bimanual basic ADLs (e.g., feeding and dressing) [30], instrumental ADLs (e.g., housework, meal preparation, and technology use)

[31], and ADLs that they found challenging without their conventional prostheses (e.g., loading a pillow into a pillowcase, hammering, donning and doffing a ring) (Fig. 9; Video S2). Improvements in ADLs are difficult to quantify, but the participants' success across a variety of different activities, including those challenging with conventional prostheses, highlights the inuitive and dextrous control afforded by the MKF.

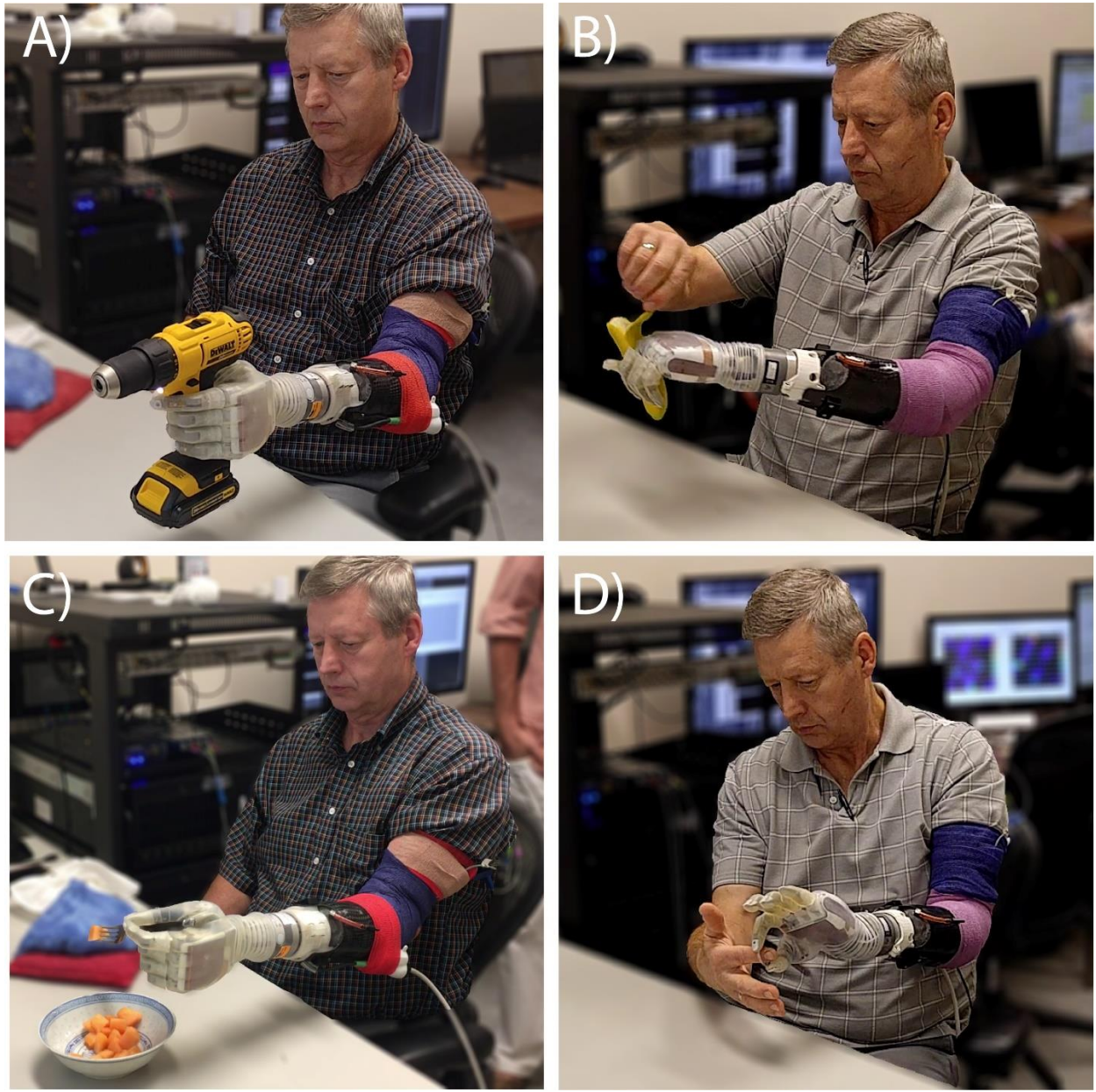

Figure 9: Amputee participants used the modified Kalman filter to actively control all six degrees of freedom of the DEKA "LUKE" Arm in order to complete various activities of daily living. Standard activities of daily living – such as using tools, feeding and dressing – were performed in the lab environment. Images show amputee participant S6: A) using a drill, B) peeling a bananna, C) eating with a fork, and D) donning/doffing a ring.

### *3.8. Computational efficiency enables portable use in a home-environment*

The above ADLs were performed in a lab environment, with the paricipant tethered to a powerful computer. An important aspect of any prosthetic control algorithm is whether or not it can be practically translated into a viable clinical product. To this end, we deployed the MKF onto the portable Nomad neural processing unit to demonstrate the feasibility of this algorithm as a clinical device.

S7 donned the prosthesis, the Nomad system, and then performed a training session in which he mimicked pre-programmed movements of the prosthesis (Video S1). In less than five minutes, the MKF was trained automatically (including auto selection of EMG channels with the Gram-Schmidt method), without any user input or experimenter intervention. S7 also performed various surpervised ADLs around his own home, focusing on ADLs that he found difficult with his conventional prosthesis (Fig. 10; Video S3). Prior complex regional pain syndrome and multi-year disuse (i.e., severe muscle atropy), as well as ongoing dystonia, made it difficult for S7 to complete ADLs with his clinically perscribed prosthesis. However, using the MKF and the LUKE arm, S7 was able to complete these activities, and he consistently demonstrated both pleasure and surprise with this newfound functionality (Video S3).

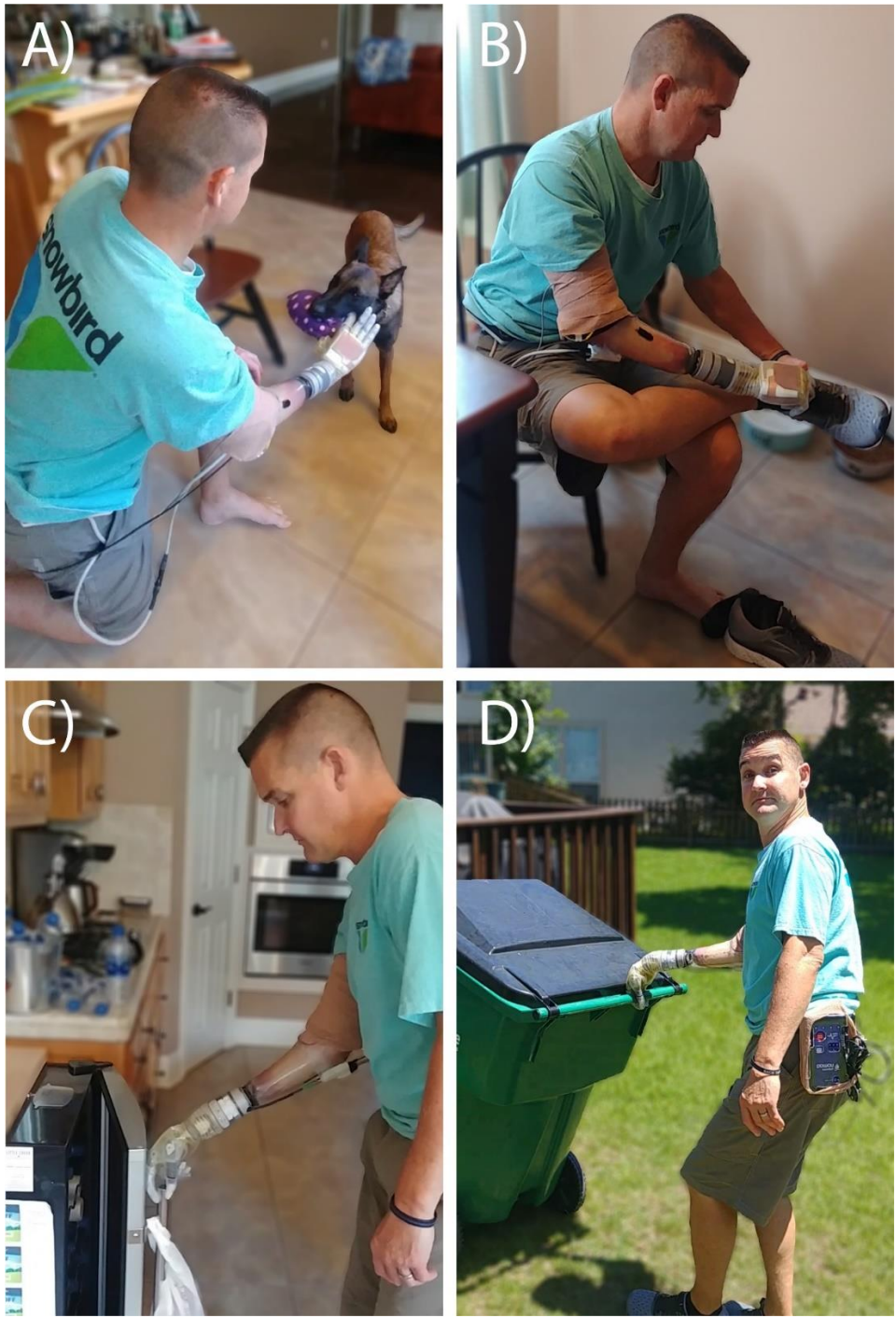

Figure 10: Amputee participant S7 performed supervised activities of daily living in a home environment using the modified Kalman filter deployed onto a portable take-home system. Images show the participant: A) playing with his dog, B) donning/doffing shoes, C) opening his refridgerator, and D) taking out the trash. S7 spontaneously reported that the portable take-home system allowed him to perform activities that were not possible with his conventional clinical myoelectric prosthesis.

## 4. Discussion:

*4.1. Summary*

In the present study, we demonstrate that modifications to the KF, namely non-uniform gains and non-zero thresholds, significantly improve prosthetic control by reducing unintended movement without degrading intended movement. Furthermore, we demonstrate that these modifications can be optimized offline and translated into functional improvements online. The MKF provides intuitive, dexterous and proportional control of high-degree-of-freedom prosthetic arms. The algorithm can be quickly trained and used for ADLs with a portable system.

*4.2. Relationship to prior work*

The MKF provides an important departure from pattern recognition control schemes for multi-articulate prosthetic hands. Although pattern recognition has been very successful at classifying hand movements from neural and/or EMG signals [32]–[35], the user is unable to control their force output. Recent studies involving sensorized prostheses have demonstrated that sensory feedback, in conjunction with proportional control, is critical for fine dexterity and manipulation of fragile objects [36]–[38].

We extend previous studies that utilized neural or neuromyoelectric signals for proportional control of 1–3-DOF prosthetic hands [36]–[38], and our prior work demonstrating up to 5-DOF neural control [12], by implementing a new technological approach—combining neural and EMG data and modifying the KF—to demonstrate proportional, 6-DOF control and practical use in ADLs.

*4.3. Benefits of the modified Kalman filter*

The simplicity of this algorithm and its ability to be implemented on a portable system are critically important. First, more advanced algorithms, such as convolutional or recurrent neural networks, require more extensive computational power (which is not always readily available on portable systems) and have yet to show substantial improvements over simpler algorithms such as the MKF. Second, the ability to retrain the algorithm, and to do so quickly (e.g., in under five minutes) allows for quick corrections in the presence of unstable neural interfaces [39] and is more practical for subject use. Third, ease of use plays a vital role in adoption of advanced neuromyoelectric prostheses. With our portable system, the participant was able to press a single button to quickly train the algorithm and then immediately begin using it in his own home.

In addition to being simple to train, the MKF also provides intuitive control over prostheses. The Gram-Schmitt channel-selection algorithm optimally selects channels that best correlate EMG/neural features to kinematics. The KF then weights them appropriately such that the same EMG/neural patterns used during training can be used again to actively control the prosthesis. All participants exhibited effective control in initial sessions, even S7 who suffered from prior CRPS and associated multi-year hand disuse. In addition, we saw no significant differences in performance due to prior experience with myoelectric prostheses (t-test between users with and without prior experience). Together, these results highlight the intuitiveness of the approach. We believe the ability to intuitively control dexterous prosthetics, despite not having prior experience, will improve ease-of-adoption, device functionality and quality of life, ultimately reducing the rate of prosthesis abandonment.

*4.4. Justification for modifications*

The KF provides an optimal estimate of a given system. Why then do ad-hoc gains and thresholds provide improvements? Indeed, with a linear system and Guassian noise, one might not expect these modifications to provide any benefit. However, neural systems are highly non-linear [40]. Additionally, patients are able to adapt to and control systems that are consistent and responsive [41]. Thresholds make the system more consistent by reducing unintended (and distracting) movements. Gains make the system more responsive by amplifying the participant's intent and reducing the effort required by the participants (e.g., less EMG activity required for maximum flexion/extension). With the user in the feedback loop to make motor adjustments, the ability to quickly correct small errors improves the overall performance, despite the mathematical inaccuracies that may be associated with ad-hoc modifications to the KF.

*4.5. Limitations and future work*

In the online task reported here, we optimized a single threshold across all six DOFs due to computational constraints and limited patient time. Our offline analyses demonstrated that the use of gains and thresholds, optimized on a DOF-by-DOF basis, improves performance compared with a standard threshold (Fig. 6).  We also demonstrated that offline performance can predict online performance. Together, these findings suggest that DOF-by-DOF modification would further improve functionality in online real-world tasks (Fig. 8). We anticipate these more complex modifications can be introduced in the near future as the computational power of portable neuromyoelectric processing units increases.

Our gains and thresholds optimization algorithm equally weighted intended and unintended movement, which may not be ideal for ADLs or accurately describe amputee preferences. This approach also inherently emphasizes unintended movement. The RMSE for movement is generally lower when the error is calculated in reference to a static kinematic value (e.g., zero) rather than a temporally varying kinematic values (e.g., flexion of a degree of freedom). Unintended movement have less RMSE than intended movement because it's easier to maintain no EMG activation than it is to maintain a fixed, non-zero EMG activation. This may be amplified by the fact that there is more training data for unintended movement than there is for intended movement (each intended movement on one DOF is associated with unintended movement on all other DOFs).

In general, thresholds reduced unintended movement but also made control of the intended DOF less accurate. Accurate intended movement may be desirable when manipulating fragile objects (to finely regulate force), whereas reduced unintended movement may be more desirable when speed is preferred over precision. Although non-uniform gains were able to counteract the negative side-effects of thresholds, ultimately the right balance between reduced unintended movement and improved intended movement may be task specific or a matter of preference.

*4.6. Broader implications*

The results presented here demonstrate the practical use of a MKF for two distinct amputee populations. Participant S7 was different from most amputees (e.g., S6, a long-term traumatic amputee) in that he underwent an elective amputation following Complex Regional Pain Syndrome. Despite being a recent amputee with severe muscle atrophy (from not using his arm for 25 years prior to the amputation) and dystonia, S7 was still able to use the MKF to complete ADLs that were not possible with his conventional prosthesis (Video S3). The combination of neuromyoelectric recordings were critically valuable to participant S7. Due to S7's severe muscle atrophy, neural recordings were used primarily for control immediately after amputation. After repeated use, and muscle strengthening and motor learning, S7's EMG features became more robust and began to mirror those of S6. Optimal modifications and performance were also similar for S6 and S7, demonstrating that the MKF can be useful for both amputee populations.

The performance achieved by S6 and S7 is due in good part to the modifications presented here. For example, the virtual hand-matching task demonstrated that S7 was critically dependent on thresholds in order to maintain a consistent hand grasp (Fig. 8). Similarly, the hold-time performance of intact participants increased by over 300% with added thresholds. Together, the ability of this approach to improve performance for two distinct amputee populations and for intact participants, using both neural and EMG data, for offline and online tasks, demonstrate the repeatability and reliability of this approach. Implementing a uniform threshold of 0.14 across all DOFs (i.e., the mean threshold utilized here) could be applied to a variety of prosthetic control algorithms; thresholds have also been utilized to improve performance of a non-linear convolutional neural network [14]. Other ad-hoc modifications, such as a latching filter, may further enhance performance and robustness for a variety of different prosthetic control algorithms [42].

The present study highlights the ability to utilize neuromyoelectric data from implanted USEAs and iEMGs for up to 14 months with S6 and up to 17 months with S7. iEMG data provided a robust measure of motor intent for both participants across the entire duration of the study. In addition, the USEAs implanted in S7 (which included improved helical wire bundles to eliminate broken wires [20]) continued to record multi-unit activity throughout the study, and the novel information provided by these neural recordings improved our ability to estimate motor intent (Fig. 4). Although not explored systematically, we speculate that the new information afforded from the neural recordings consists of efferent motor fibers or afferent proprioceptive fibers associated with intrinsic hand muscles that cannot be targeted with iEMGs after amputation. This work demonstrates that neuromyoelectric devices, such as the USEAs and iEMGs utilized in this study, can provide intuitive prosthetic control. Future improvements to these implantable devices, such as wireless communication, will play a critical role in translating the benefits of the MKF into a clinical device to better serve amputees.

## 5. CONCLUSION:

Our hands are critically important in daily life, often dicating what we can do and influencing who we are. Quality of life for upper-limb amputees is often limited by ineffective control of prostheses [4]. We demonstrated that neural and muscle activity recorded from the residual forearm can be used to provide inuititive, proportional, and high-DOF prosthetic control. Alongside novel sensory restoration strategies, these improvements in prosthesis dexterity will likely enhance prostheses embodiment [13], [43], reduce phantom pain [13], [44], and enable complex sensorimotor capabilities such as stereognosis. The results of this study set the stage for exploring the benefits of intuitive and proportional control of multi-articulate prostheses in take-home trials. Future work will utilize the portable take-home system introduced here to systematically compare the performance of the MKF with clinically available control strategies.

**ACKNOWLEDGEMENTS:**

We thank the amputee participants involved in this study for their willingness to contribute more than a year of their lifes to improve science and the lives of future amputees. We acknowledge Dr. Douglas Hutchinson and Dr. Christopher Duncan for their clinical support throughout these studies. We also thank Dr. David Kluger, Dr. Suzanne Wendelken, and Michael Paskett for assisting with experiments and Dr. David Warren for revising early drafts of this manuscript. Photos of devices courtesy of Dr. Loren Reith and the University of Utah Department of Veterans Affairs.

This work was sponsored by the Hand Proprioception and Touch Interfaces (HAPTIX) program administered by the Biological Technologies Office (BTO) of the Defense Advanced Research Projects Agency (DARPA) through the Space and Naval Warfare Systems Center, Contract No. N66001-15-C-4017. Additional sponsorship was provided by the National Science Foundation (NSF) through grant No. NSF ECCS-1533649, the NSF Graduate Research Fellowship Program Award No. 1747505, and the University of Utah Department of Veterans Affairs. The content herein represents the views of the authors, not their employers' or funding sponsors'.

**SUPPLEMENTAL MATERIAL:**

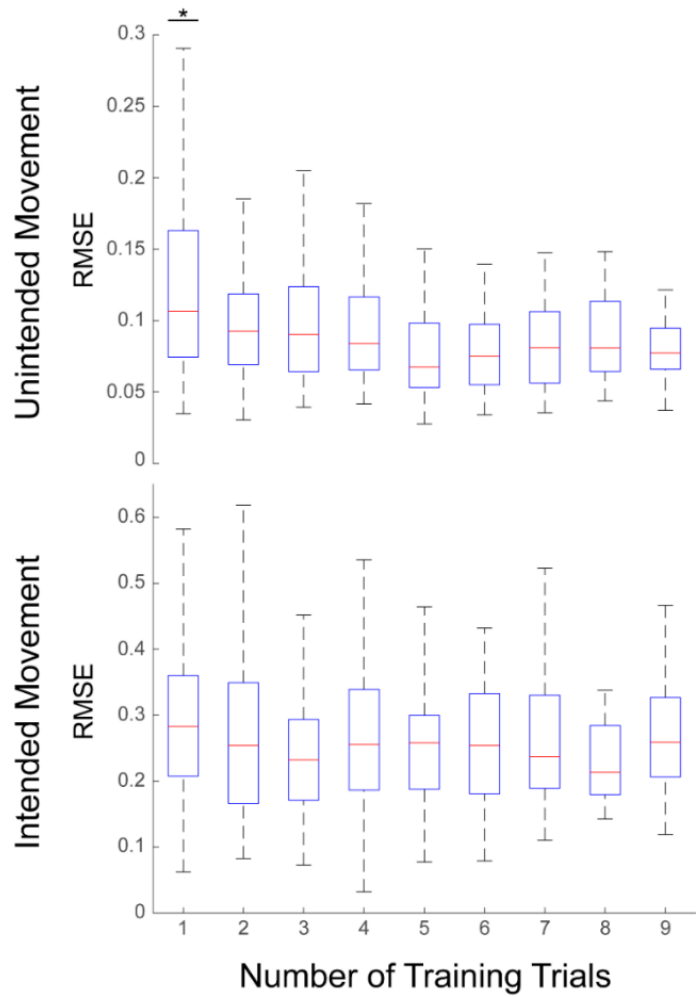

Figure S1: Performance of an unmodified Kalman filter as a function of training trials. There were no significant differences between the number of trials for the root mean squared error (RMSE) of intended movement. However, the RMSE of unintended movement was significantly worse when trained on only one trial. Data are from 66 datasets with 10 trials each from amputee participant S6. The Kalman filter was trained with 1–9 trial(s) and tested on the remaining 9–1 trial(s). Subsequent datasets with amputee participant S7 used four trials to minimize training time while still ensuring at least two training

trials for offline analyses (when half the data is partitioned for testing). * $p < 0.01$ for all pairwise comparisons.